\documentclass[aps,prl,reprint,floatfix,showpacs]{revtex4-1}
\usepackage{amssymb,amsmath}
\usepackage{graphicx}

\usepackage[scanall]{psfrag}

\usepackage{color}

\newcommand{\psLabelSize}{1.3}
\newcommand{\psSymbolSize}{0.8}
\newcommand{\psSweepSize}{0.9}
\newcommand{\abs}[1]{\left| #1 \right|}

\graphicspath{{./Figs/fraggedFigs/}}

\begin{document}

\title{Designing Isotropic Interactions for Self-Assembly of Complex Lattices}%

\author{E. Edlund}%
\author{O. Lindgren}
\author{M. \surname{Nilsson Jacobi}}%
 \email{mjacobi@chalmers.se} 
\affiliation{Complex Systems Group, Department of  Energy and Environment, Chalmers University of Technology, SE-41296 G\"oteborg, Sweden}
\date{\today}%

\begin{abstract}
We present a direct method  for solving the inverse problem of designing isotropic potentials that cause self-assembly into target lattices. Each potential is constructed  by matching its energy spectrum to the reciprocal representation of the lattice to guarantee that the desired structure is a ground state. We use the method to self-assemble complex lattices not previously achieved with isotropic potentials, such as a snub square tiling and the kagome lattice. The latter is especially interesting because it provides the crucial geometric frustration in several proposed spin liquids.

\end{abstract}


\maketitle


Self-assembly~\cite{whitesides_self-assembly_2002}, the process through which simple entities spontaneously form complex structures, is a phenomenon of fundamental interest and a compelling tool for future nanoscale fabrication. As technology is pushed to ever smaller length scales, direct fabrication becomes increasingly challenging and material design may turn towards tailoring interactions which causes constituents to order into the desired patterns without directed external influence. This vision has been the subject of intense experimental work during the last decades but there is still a relative lack of theoretical results beyond pure simulation based work.

What theoretical work exists can be understood as trying to solve an ''inverse'' statistical mechanics problem, finding interactions that causes self-assembly into a target structure or configurations with desired properties. Various schemes for approaching this problem have been proposed~\cite{torquato_inverse_2009, cohn_algorithmic_2009} and several isotropic potentials self-assembling into low-coordinated lattices have been reported, e.g.\ square, honeycomb, simple cubic, and diamond lattices~\cite{rechtsman_optimized_2005, *rechtsman_self-assembly_2006, *rechtsman_synthetic_2007}. The focus is on isotropic interactions, partly due to theoretical tractability but mainly motivated by the recent years' remarkable improvement in understanding and design capabilities of such interactions in especially colloidal and nano systems~\cite{likos_effective_2001,min_role_2008}. By combining and tuning forces including, but not limited to, van der Waals, electrostatic, magnetic, and steric forces as well as forces of purely entropic origin we can today construct complex effective interactions. With such a profusion of possibilities, the forward statical mechanics' prescription of sweeping the parameter space to give  a complete picture of the possible structures becomes  computationally infeasible, and the inverse formulation increasingly appealing.

In this Letter we present a new method for designing potentials for targeted self-assembly  based on an energy spectrum approach. The method is direct in contrast to iterative or optimization approaches more commonly used to solve inverse problems. We show how interactions that cause particles to organize into the kagome lattice can be obtained. The kagome lattice is a much sought after topology due to its highly frustrated geometry~\cite{ramirez_strongly_1994, greedan_geometrically_2001}, possibly resulting in spin liquid ground states in several materials~\cite{helton_spin_2007,okamoto_vesignieite_2009} as well as in the corresponding Heisenberg model~\cite{yan_spin-liquid_2011, *[{but see also }]singh_ground_2007, *evenbly_frustrated_2010}. The self-assembly of nanoscale~\cite{moulton_crystal_2002,paul_organically_2002} and colloidal~\cite{chen_directed_2011} kagome lattices through directed interactions has caused considerable interest but whether this target structure is achievable via isotropic potentials have until now been an open question (cf.\ the negative result in~\cite{grivopoulos_no_2009}). To demonstrate the generality of our method, we also apply it to the honeycomb lattice and a snub square tiling~\cite{grunbaum_tilings_1986}.

\begin{figure}[b]
\psfrag{a}[c][c][\psLabelSize][0]{\color[rgb]{1,1,1}\bf{a}}
\psfrag{b}[c][c][\psLabelSize][0]{\color[rgb]{1,1,1}\bf{b}}
\includegraphics[width=0.45 \textwidth]{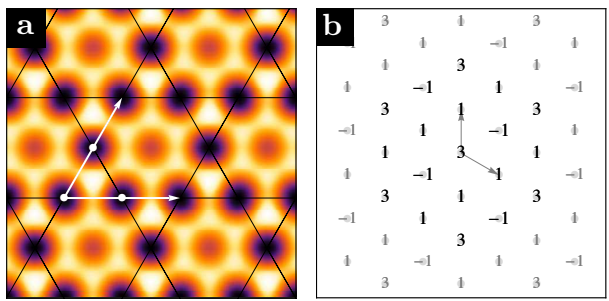}
\caption{ \label{kagLowRes} 
The kagome lattice in real space (a) as described by its first few reciprocal lattice vectors (b). Structure factors for the included reciprocal vectors are shown in black and primitive lattice vectors ${\bf k}_p$ as gray arrows. The kagome lattice (illustrated in black, with basis and lattice vectors in white) is indeed determined by this limited part of the reciprocal lattice.
}
\end{figure}

We consider self-assembly governed by energy minimization where an interaction is designed with certain features so that the target structure is a stable low energy state, which should typically be reached when the system self-assembles from random initial configurations. Our method differs from previous ones in that we use a reciprocal representation of the lattice to construct a theoretical potential in a way that guarantees the target structure to be the unique ground state. The potential is then systematically simplified into a more realizable form. In general a lattice structure is defined by the small $| {\bf k} |$ region of the Fourier spectrum while the large $| {\bf k} | $ region only contributes with higher resolution of the distribution defining the particles. The kagome lattice for instance is clearly resolved (Fig.~\ref{kagLowRes}) when expressed as the sum of the Fourier modes of the (weighted) reciprocal lattice vectors up to length $2 |{\bf k}_p|$, where ${\bf k}_p$ is a primitive reciprocal lattice vector. This makes the reciprocal perspective a natural approach for targeted self-assembly of lattice structures, as opposed to real-space methods focusing on the pair correlation function~\cite{rechtsman_optimized_2005, *rechtsman_self-assembly_2006, *rechtsman_synthetic_2007}.

In two dimensions the potential energy for a configuration with particle density $\rho ( {\bf r} )$ can be expressed in reciprocal space as 
\begin{eqnarray} \label{fourierEnergy}
	E & = &  \int \! \mathrm{d} {\bf r}_1 \mathrm{d} {\bf r}_2 \, \rho ({\bf r}_1 ) V ( | {\bf r} _1 -  {\bf r} _2 | ) \rho ({\bf r}_2 ) \nonumber \\
	& = &  \int \! \mathrm{d}k \,\abs{\hat{\rho}(k)}^2 \widehat{V}(k) , 
 \end{eqnarray} 
 where $|\hat{\rho} (k)|^2  \mathrm{d}k= \int _{ {\bf k} = k } \mathrm{d} {\bf k} \, |\hat{\rho} ( {\bf k} )|^2$, $\hat{\rho} ( {\bf k} )$ is the standard Fourier transform of $\rho ( {\bf r} )$, and $\widehat{V}(k)$ is the radial Fourier (Hankel) transform of $V(r)$~\cite{folland_fourier_2009},
\begin{equation}
	\widehat{V}(k) = 2 \pi \int \! r \, \mathrm{d} r \, V(r) J_0 (r k).
	\label{Vk}
\end{equation}
If $V( {\bf r} _1 , {\bf r} _2  ) =   V(|  {\bf r} _1 - {\bf r} _2 | )$ is interpreted as a linear operator, then Eq~\eqref{fourierEnergy} defines a diagonalization of $V( {\bf r} _1 , {\bf r} _2  )$. We therefore refer to $\widehat{V}(k)$ as the energy spectrum of the potential~\cite{edlund_universality_2010}.

We use the power spectrum of the target lattice $ \abs{\hat{\rho}(k)}$ and Eq.~\eqref{fourierEnergy} to construct a potential that guarantees that the target is a ground state in the following way. Consider a potential $V(r)$ whose energy spectrum is chosen positive, $\widehat{V}(k)\geq0$, and has zeros only at those points coinciding with the reciprocal lattice points ${\bf G}_i$ of the Bravais lattice describing the target lattice, $\widehat{V}(\abs{{\bf G}_i})=0$ $\forall i$. This design guarantees that the target lattice is a ground state with $E=0$. However, any lattice with reciprocal points $\subseteq {\bf G} _i$, i.e.\ with arbitrary basis (as well as reciprocal lattices with opposite chirality), is also a ground state. We break this degeneracy in favor of the target lattice by introducing small perturbations of the energy spectrum at its zeros and use the fact that there are amplitude differences in the basis dependent $\hat{\rho}({\bf G}_i)$.

To understand how perturbations affect the energy levels there are two characteristics of $\hat{\rho}({\bf k})$ we need to consider. For a closed system with $n$ particles $\hat{\rho}({\bf 0})=n$ so $\hat{V}(0)$ is irrelevant to the ground state configuration. More importantly, the limited mass provides an upper limit for $\abs{\hat{\rho}({\bf k})}=\abs{\Sigma_{i=1}^n e^{i {\bf k} \cdot {\bf r}_i}}\leq n$, with equality for certain ${\bf k}$ if ${\bf r}_i$ coincides with the resonance points of the modulation. It is this property we use to break the degeneracy.

\begin{figure*}[tb]
\psfrag{a}[c][c][\psLabelSize][0]{\color[rgb]{1,1,1}\bf{a}}
\psfrag{b}[c][c][\psLabelSize][0]{\color[rgb]{1,1,1}\bf{b}}
\psfrag{c}[c][c][\psLabelSize][0]{\color[rgb]{1,1,1}\bf{c}}
\psfrag{d}[c][c][\psLabelSize][0]{\color[rgb]{1,1,1}\bf{d}}
\psfrag{E}[c][c][\psSymbolSize][0]{$\widehat{V}(k)$}
\psfrag{k}[c][c][\psSymbolSize][0]{$k$}
\psfrag{V}[c][c][\psSymbolSize][0]{$V(r)$}
\psfrag{r}[c][c][\psSymbolSize][0]{$r$}
\psfrag{S}[c][c][\psSymbolSize][0]{\color[rgb]{1,0,0}$\abs{\hat{\rho}(k)}^2 \mathrm{d}k$}

\psfrag{t}[t][c][1][0]{$\frac{4\pi}{\sqrt{3}}$}
\psfrag{M}[c][c][\psSweepSize][0]{$t = 0$ sweeps}
\psfrag{N}[c][c][\psSweepSize][0]{$t = 2\cdot 10^4$ sweeps}
\psfrag{O}[c][c][\psSweepSize][0]{$t = 10^5$ sweeps}
\psfrag{P}[c][c][\psSweepSize][0]{$t =  10^6$ sweeps}

\includegraphics[width=0.95 \textwidth]{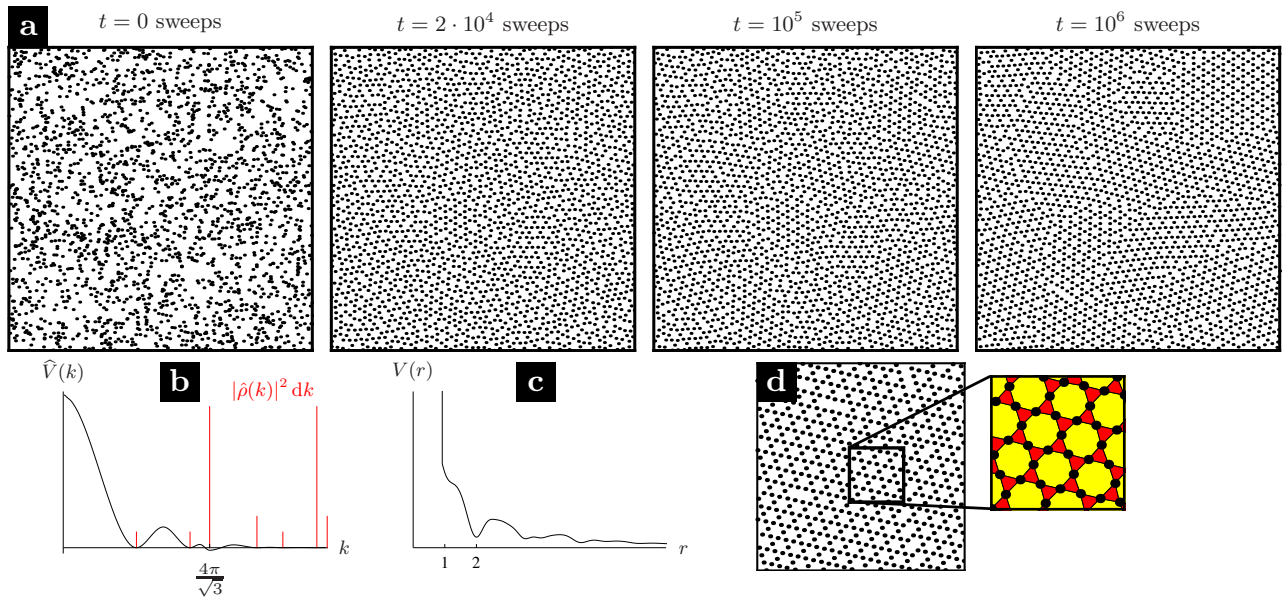}
\caption{ \label{latticeFig} 
A spectral method for targeted self-assembly. By designing an energy spectrum $\widehat{V} (k)$ (b) with minima at the peaks of ${\color[rgb]{1,0,0} \hat{\rho}}$ and taking its Fourier transform we arrive at a potential $V(r)$ (c) that causes particles to self-assemble into a target lattice. (a) Time evolution of a self-assembling kagome lattice. The number of grains diminishes by random walk of their boundaries, a process faster for smaller systems (d) but where some local defects are still present as the dimensions of the system is not fine tuned to fit the lattice.}
\end{figure*}

We now apply this design scheme to the kagome lattice. Consider an energy spectrum with zeros corresponding to the hexagonal Bravais lattice with primitive vectors $\left\{(2,0),(1,\sqrt{3})\right\}$, which encompasses the kagome lattice (basis $\left\{(0,0),(1,0),(1/2,\sqrt{3}/2)\right\}$). The maximum contribution to the Fourier transform $\hat{\rho}({\bf k})=\sum_{{\bf r}_j\in \mathrm{basis}} e^{i \pi {\bf k} {\bf r}_j}$ at the reciprocal points $\abs{{\bf G}_i}=\frac{4 \pi}{\sqrt{3}}$ is given by ${\bf r}_j\in \left\{(0,0),(0,1),(1/2,\sqrt{3}/2),(3/2,\sqrt{3}/2)\right\}$. With a particle density of approximately three particles per primitive cell any subset of three of the points ${\bf r}_j$ is a basis for the kagome lattice. Including a hard core repulsion  makes ${\bf r}_j={\bf r}_l$, $j\neq l$, unfeasible without energetically benefiting any configuration nor affecting the kagome lattice. Therefore a negative perturbation of $\widehat{V}(k)$ at $k=\frac{4 \pi}{\sqrt{3}}$ breaks the degeneracy in favor of the kagome lattice which becomes the unique ground state. An energy spectrum constructed by following these prescriptions is shown in Fig.~\ref{latticeFig}.

A typical energy spectrum fulfilling the above constraints results in a highly complex potential, not realizable in experimental systems. However, the properties of the energy spectrum that select the ground state are far from a complete specification. This has dual consequences in that, on one hand, there is considerable leeway in varying experimental realizations for a target system but, on the other hand, the prescriptions leaves a fairly large space of possible designs to be searched. While having no effect on the ground states, these choices  can greatly affect the realizability of the potential as well as the extent to which a system of particles interacting with the resulting potential can achieve the ground state under realistic dynamics. Reasonable requirements on the designed potentials include smoothness and restrictions on their range. To understand how this can be achieved we note that the Fourier relation in Eq.~\eqref{Vk} implies that there exist a version of Heisenberg's uncertainty principle relating the smoothness of the potential in real space to a screening in reciprocal space, and vice versa. Several general principles on how the power spectrum should be designed follow from this observation. Restricting the power spectrum to the low ${\bf k}$ region gives smoother potentials, evident in Fig.~\ref{latticeFig}. Notice how the first few minima of the power spectrum is sufficient to determine the lattice periodicity. A screening of the power spectrum provides further smoothing of the potential %
\footnote{
For a more formal treatment of the screening-smoothing duality, we use the fact that $J_0$ is an eigenfunction of the radial Laplace operator $\Delta _k J_0 (k r) = -r^2 J_0 (k r)$ to write the relation 
\[
 \Psi [ \Delta _k ]  \hat{V}(k) = \int \! r\mathrm{d} r \, V(r) \Psi (-r^2) J_0(k r)
\]
with any function $\Psi$ for which the integral converges. By choosing e.g.\ $\Psi [ X ] = e^{a X}$ one see that a gaussian smoothing of the spectrum corresponds to a screening of the potential and vice versa.
}. %
Conversely, to get a short ranged potential the spectrum should be smooth; in Fig.~\ref{latticeFig} this is achieved using third orders splines as interpolations between the extreme points.


To test our method we perform Monte Carlo simulations with the constructed interaction potentials to confirm that they reliably and in a positive density range self-assemble into their respective target lattices. All simulations starts in a random disordered state and we consider only local moves of the particles in the Monte Carlo updates. We do not fine-tune the dimensions of the simulated systems, therefore some deformations of the lattices are necessary for it to fit. The simulations confirm that the designed potentials are robust to such complications.
 
 Fig.~\ref{latticeFig}a illustrates the relaxation process for the physically most interesting structure, the kagome lattice. The snapshots of the time evolution show how the lattice assemble locally into grains with boundaries that diffuse and slowly annihilate. The system size in Fig.~\ref{latticeFig}a is chosen large enough to illustrate this process but makes complete relaxation too computationally demanding. In Fig.~\ref{latticeFig}d we show the global ground state for a smaller system with the same potential. It should be noted that the kagome lattice is significantly more complex than lattices previously achieved through targeted self-assembly with isotropic interactions. 
 
\begin{figure}[tb]
\psfrag{a}[c][c][\psLabelSize][0]{\color[rgb]{1,1,1}\bf{a}}
\psfrag{b}[c][c][\psLabelSize][0]{\color[rgb]{1,1,1}\bf{b}}
\psfrag{V}[c][c][\psSymbolSize][0]{$V(r)$}
\psfrag{r}[c][c][\psSymbolSize][0]{$r$}

\includegraphics[width=0.49 \textwidth]{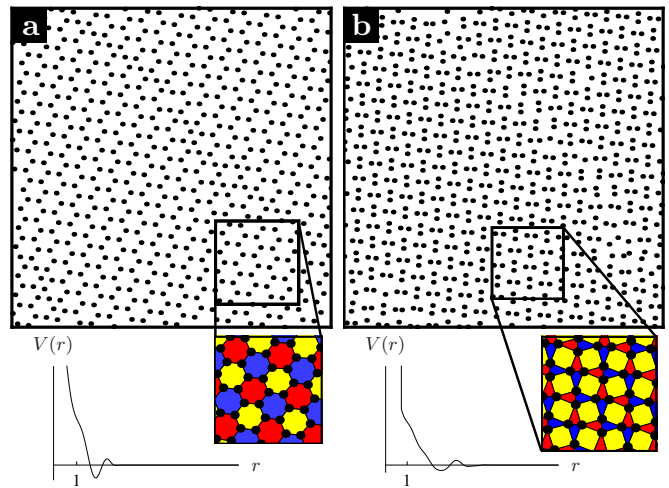}
\caption{ \label{confFig} 
Isotropic potentials that causes particles to self-assemble into honeycomb (a) and snub square (b) lattices.
}
\end{figure}

 To demonstrate the range of our method, we also assemble a snub square tiling with isosceles triangles~\cite{grunbaum_tilings_1986}, shown in Fig.~\ref{confFig}. This structure is even more complex than the kagome lattice since it has a square Bravais lattice and requires more perturbations to single out the correct basis. We use perturbations at four points in the energy spectrum, three positive perturbations where $\hat{\rho}({\bf G}_i)=0$ and one negative where  $\hat{\rho}({\bf G}_i)=n$.
 
For completeness we also show how our method applies to the previously self-assembled honeycomb lattice~\cite{rechtsman_optimized_2005}. It is significantly simpler, but physically interesting especially in its graphene incarnation~\cite{castro_neto_electronic_2009}. Since the power spectrum has fewer features ($\hat{\rho}({\bf G}_i)\in\left\{2,1\right\}$), even a relatively simple potential can exhibit the crucial properties assuring that the honeycomb lattice will self-assemble.


All examples above utilizes key features of the lattice in reciprocal space, namely that  the power spectrum reaches its theoretical maximal limit at reciprocal lattice sites $|{\bf G}_i|=k$ for some $k$. If this criterion is not fulfilled,  as is the case for several Archimedean tilings like the equilateral snub square, one cannot reduce the group of possible basis configurations to a finite number (discrete positions) and the choice of perturbations is not as straight forward. However, allowing for competing perturbations expands the group of achievable lattices. Since the most distinct features of the energy spectrum is those determining the Bravais lattice, relevant for the interactions between particles of different unit cells, the method is simpler to apply to lattices with small unit cells. 

In general, proving that a certain configuration is the ground state of a given potential is a very hard problem. In fact, the exact nature of the ground state is not rigorously known even for simple interactions such as the Lennard-Jones potential~\cite{Schachinger}. In this Letter we have described a direct method to design potentials for targeted self-assembly of lattices, a problem usually approached using iterative methods involving repeated relaxations of the system~\cite{torquato_inverse_2009, cohn_algorithmic_2009}. From our construction follows the somewhat counterintuitive observation that it is actually simpler to find a potential with a given configuration as a ground state than to determine the ground state(s) of a given potential.

To conclude, we have demonstrated that there exist isotropic potentials that produce complex lattices such as the kagome. Previous work addressing targeted lattice formation have focused on the structure of the potential in real space, e.g.~\cite{torquato_inverse_2009, cohn_algorithmic_2009}, whereas our approach focuses on the energy spectrum of the potential. We believe our method to be better suited for targeted self-assembly of lattices since their translational invariance  is more naturally expressed in reciprocal space. 

The ability to design interactions that self-assemble into complex lattices is interesting in itself, but this remains a theoretical observation if the potentials are too complicated to be feasible except in computer simulations. By analyzing how smoothing and screening affect the energy spectrum we can systematically simplify the potentials and thereby design interactions that should be possible to implemented in experimental systems.

\begin{acknowledgments}
OL and MNJ acknowledge support from the SuMo Biomaterials center of excellence.
\end{acknowledgments}

\end{document}